\documentclass{article}

\usepackage{microtype}
\usepackage{graphicx}
\usepackage{subcaption}
\usepackage{booktabs} 

\usepackage{hyperref}
\usepackage{xurl}

\usepackage[preprint]{icml2026}

\usepackage{amsmath}
\usepackage{amssymb}
\usepackage{mathtools}
\usepackage{amsthm}

\usepackage[capitalize,noabbrev]{cleveref}

\theoremstyle{plain}

\theoremstyle{definition}

\theoremstyle{remark}

\usepackage{listings}
\usepackage{xcolor}

\usepackage[textsize=tiny]{todonotes}

\icmltitlerunning{Unveiling the Potential of Quantization with MXFP4: Strategies for Quantization Error Reduction}

\begin{document}

\twocolumn[
  \icmltitle{Unveiling the Potential of Quantization with MXFP4:\\Strategies for Quantization Error Reduction}
  


  \icmlsetsymbol{equal}{*}

  \begin{icmlauthorlist}
    \icmlauthor{Jatin Chhugani}{equal,comp}
    \icmlauthor{Geonhwa Jeong}{equal,comp}
    \icmlauthor{Bor-Yiing Su}{comp}
    \icmlauthor{Yunjie Pan}{comp}
    \icmlauthor{Hanmei Yang}{comp}
    \icmlauthor{Aayush Ankit}{comp}
    \icmlauthor{Jiecao Yu}{comp}
    \icmlauthor{Summer Deng}{comp}
    \icmlauthor{Yunqing Chen}{comp}
    \icmlauthor{Nadathur Satish}{comp}
    \icmlauthor{Changkyu Kim}{comp}
  \end{icmlauthorlist}

  \icmlaffiliation{comp}{Meta Platforms, Inc., Menlo Park, USA}

  \icmlcorrespondingauthor{Jatin Chhugani}{jatinch@meta.com}

  \icmlkeywords{Machine Learning, ICML}

  \vskip 0.3in
]



%
\printAffiliationsAndNotice{\icmlEqualContribution}

\begin{abstract}
Large Language Models (LLMs) have intensified the need for low-precision formats that enable efficient, large-scale inference. The Open Compute Project (OCP) Microscaling (MX) standard is attractive due to its favorable hardware efficiency, but its 4-bit variant (MXFP4) lags behind NVIDIA’s NVFP4 in accuracy, limiting adoption.
We introduce two software-only techniques, \emph{Overflow-Aware Scaling} (OAS) and \emph{Macro Block Scaling} (MBS), that improve MXFP4 quantization fidelity without requiring hardware changes. OAS reduces overall errors by increasing effective dynamic range under power-of-two block scaling, while MBS allocates higher-precision scaling at a coarser granularity to better preserve outliers.

Across multiple LLMs and standard downstream benchmarks, OAS and MBS reduce the end-to-end accuracy gap between MXFP4 and NVFP4 from about 10\% to below 1\% on average, while incurring modest GEMM overhead (6.2\% on average). These results re-establish MXFP4 as a practical alternative to NVFP4, enabling near-NVFP4 accuracy while retaining MX’s hardware-efficiency advantages (e.g., 12\% relative area savings in tensor cores).
\end{abstract}
\section{Introduction}
\label{sec:intro}

Large Language Models (LLMs) are rapidly transforming the landscape of artificial intelligence, driving breakthroughs across a wide range of applications. As the demand for higher performance continues to grow, researchers are scaling these models to unprecedented sizes. However, this scaling comes with significant computational and resource challenges, making efficiency a critical concern.

Quantization has emerged as a promising solution to address these challenges, enabling more efficient deployment of LLMs by reducing the precision of model parameters. Among various quantization formats, the microscaling format (MX) has gained traction and is becoming a standard, largely due to its adoption and promotion by multiple companies through the Open Compute Project (OCP)~\cite{ocp_microscaling_v1}. The MX proposal includes a family of formats, ranging from 8-bit and 6-bit down to 4-bit formats. While there have been successful demonstrations of the 8-bit and 6-bit formats~\cite{mx_arxiv, mx8_training}, preserving model quality with the MXFP4 format remains a significant challenge~\cite{mr_gptq, nvfp4_training, quartet_fp4}.

Consequently, NVIDIA has proposed a new 4-bit format, NVFP4~\cite{nvfp4_format} which has higher representation fidelity than the MXFP4 format. A few studies also showed that NVFP4 preserves the model quality better~\cite{nvfp4_training, nvfp4_kv_cache, mr_gptq, abdelfattah2025_razer_blog,fp4alltheway}. This fidelity gap poses a significant barrier to the widespread adoption of MXFP4 in scenarios where model performance is paramount. However, supporting the NVFP4 format incurs extra area and energy overheads to the hardware design. We perform a detailed analysis comparing the MXFP4 and NVFP4 in terms of representation fidelity and hardware costs. Building on these insights, we propose strategies to push the limits of MXFP4 quantization, achieving improved accuracy without requiring any hardware changes\footnote{While this work focuses on MXFP4, our proposed methods are generalizable to other MX formats, such as MXFP6 and MXFP8.}.
In this paper, our primary contributions are:

\begin{enumerate}
\item We identify the two primary sources of MXFP4’s accuracy gap relative to NVFP4, coarser block granularity and power-of-two scaling precision, and quantify their fidelity and hardware-area trade-offs.
  \item We propose \emph{Overflow-Aware Scaling} (OAS) and \emph{Macro Block Scaling} (MBS), two SW techniques that improve MXFP4 representation fidelity without requiring hardware modifications, making them applicable to MXFP4-compatible devices.
  \item We demonstrate that enhanced MXFP4 achieves near-NVFP4 fidelity (within 1 dB QSNR) and downstream accuracy (within 1\% on average), with modest GEMM overhead (6.2\% on average), thereby unlocking MXFP4’s hardware-efficiency benefits.
\end{enumerate}
\section{Background}
\label{sec:background}

\subsection{Transformer}
\begin{figure}[t]
\centerline{\includegraphics[width=0.95\columnwidth]{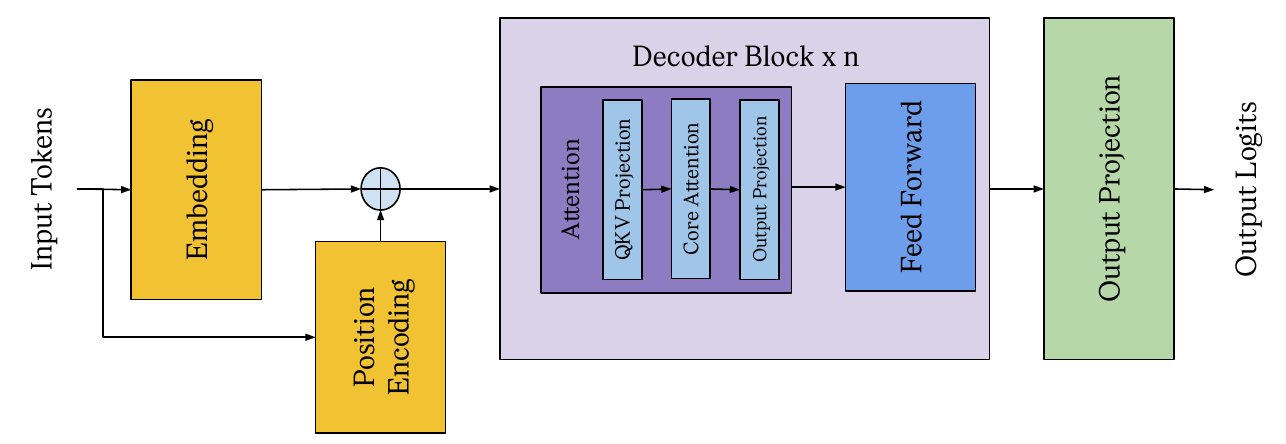}}
\caption{
Modern LLM Model Architecture.
}
\label{fig:llm_arch}
\end{figure}

Modern large language models (LLMs), including Llama 3~\cite{grattafiori2024llama3herdmodels}, Llama 4~\cite{llama4_maverick_17b}, Qwen~\cite{qwen3_8b}, DeepSeek~\cite{deepseek_r1}, and GPT-OSS~\cite{gpt_oss}, are built upon the transformer architecture. \autoref{fig:llm_arch} illustrates the decoder-only transformer used by these models. The dominant computation arises from linear layers in the QKV projections, output projection, and feed-forward network (FFN). In this work, we focus on improving the efficiency of these linear layers through weight and activation quantization.

\subsection{Quantization with FP4}
\begin{figure}[t]
\centerline{\includegraphics[width=0.95\columnwidth]{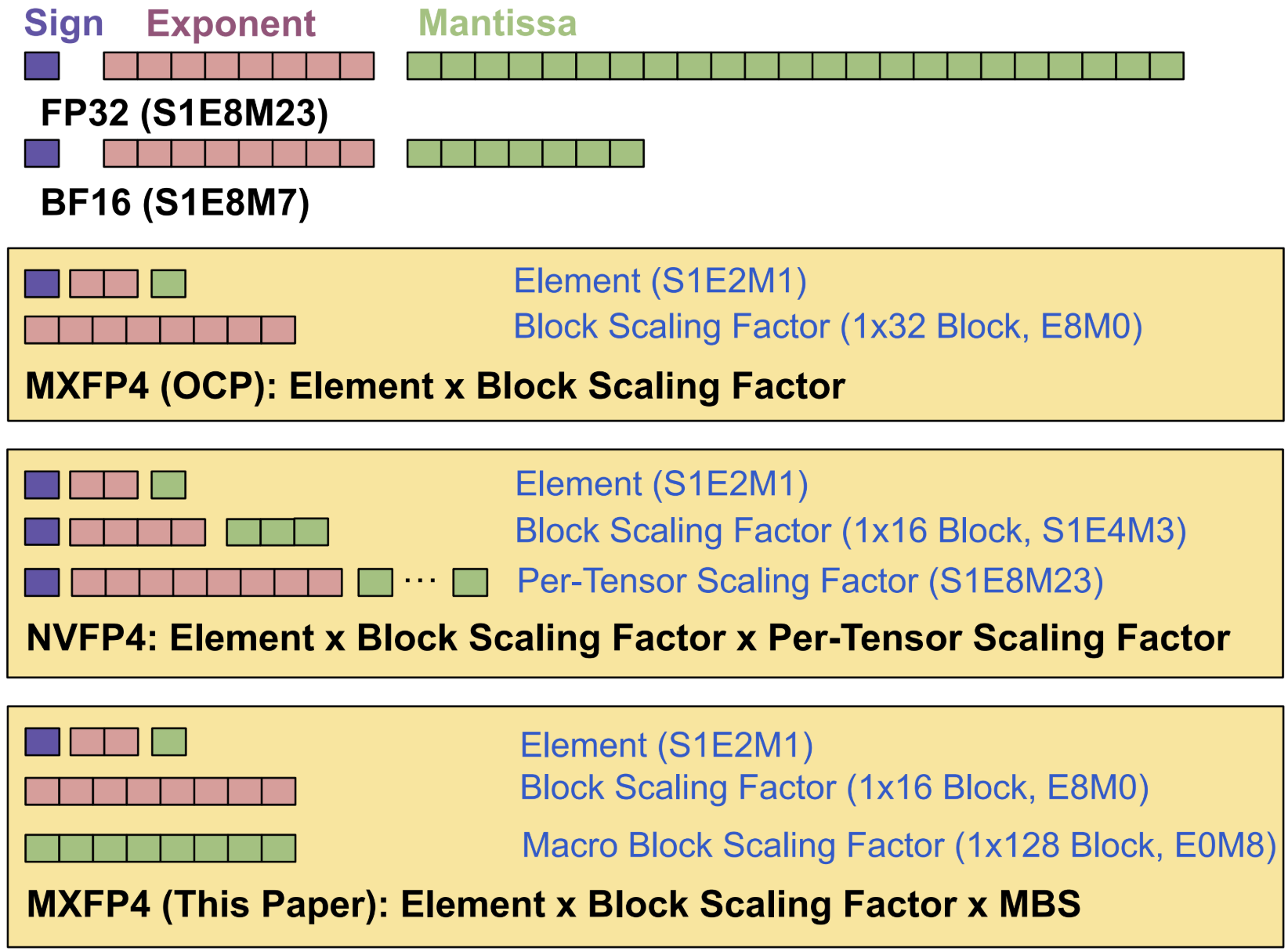}}
\caption{
Comparison of different FP4 formats for quantization. 
}
\label{fig:mxfp4_nvfp4}
\end{figure}
Quantization reduces the precision of weights and activations to improve the efficiency of LLM inference. Common schemes vary in granularity (e.g., per-tensor, per-channel, or per-group) and scaling strategy (symmetric or asymmetric)~\cite{mor2025,paulius_fp8,smooth_quant,torchao}. Furthermore, quantization-aware training (QAT) or post-training quantization (PTQ) techniques can be further employed to optimize model performance under reduced precision. 
Recent low-bit formats, such as MXFP4 and NVFP4, further explore this trade-off by enabling aggressive precision reduction while aiming to preserve model accuracy. These two formats represent prominent 4-bit quantization approaches for LLM deployment~\cite{ocp_microscaling_v1, nvfp4_format}. NVFP4, developed by NVIDIA, is widely adopted due to its strong accuracy and compatibility with existing hardware, whereas MXFP4, standardized by the Open Compute Project (OCP), is gaining attention for its improved HW efficiency. 
The key differences between these formats lie in their numerical encoding schemes and hardware implementation requirements, as shown in \autoref{fig:mxfp4_nvfp4}. Standard floating-point (FP) numbers comprise one sign bit, $E$ bits for the exponent, and $M$ bits for the mantissa.

The OCP MXFP4 format comprises two components: 4-bit data elements in E2M1 and a shared E8M0 block scale applied to every 32 elements. 
In contrast, NVFP4 is composed of three components: the 4-bit data elements in E2M1, a shared E4M3 FP8 block scale applied to every 16 elements, and a per-tensor scaling factor to mitigate range limitations. 
While NVFP4 generally provides higher representation fidelity, MXFP4 offers substantial resource savings, making it attractive for large-scale and energy-efficient deployments.

Our proposed advanced MXFP4 format is similarly defined as a combination of three components to enhance fidelity. The details of this format are provided in \autoref{subsec:mbs}. 

\subsection{HW for MXFP4 GEMM and NVFP4 GEMM}
\begin{figure}[t]
\centerline{\includegraphics[width=0.75\columnwidth]{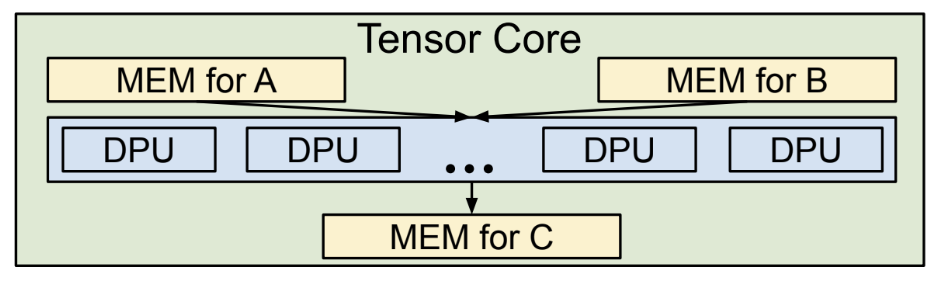}}
\caption{Hardware architecture of Tensor Core.
}
\label{fig:tc_hw}
\end{figure}

General Matrix Multiplication (GEMM) operations are central to LLM inference. 
Hardware support for GEMM using MXFP4 and NVFP4 formats varies based on the underlying architecture. 
\autoref{fig:tc_hw} shows the hardware implementation of a typical tensor core \cite{stc2019micro, nervana, darvish2023shared} which can support MXFP4/NVFP4 as input data types.
Memory for A and B store the input operand matrices and Memory for C stores the partial sums or final output.
Multiple Dot Product Units (DPU) instantiated based on Performance, Power, Area constraints, perform a HW tile size for the target GEMM operation. 
We present our detailed comparison on HW overhead of the different FP4 formats in \autoref{subsec:hw_overhead}.
\section{Understanding NVFP4 vs. MXFP4}
\label{sec:analysis}
\subsection{Analysis Methodology} In this paper, we evaluate representational fidelity using the Quantization Signal-to-Noise Ratio (QSNR), measured in decibels (dB). While different metrics can be used, we adopt QSNR~\cite{darvish2023shared} since it exhibits strong correlation with end-to-end metrics of inference quality and is also used by other works~\cite{darvish2023shared, mr_gptq} to derive a first-order analysis of the format choices.
We compute QSNR at two granularities: for individual (input) tensors and for post-operation results (e.g., the output of MatMul). This allows us to normalize the Mean Squared Error (MSE) of the quantized tensors relative to their high-precision counterparts. A higher value of QSNR signifies a lower error due to quantization, and hence higher fidelity.

Formally, let $A^{\text{BF16}}$ denote the tensor in the original high precision  and $A^Q$ denote the quantized tensor. The QSNR is defined as the logarithmic ratio of the reference signal power to the quantization noise:

\begin{align}
    & \text{QSNR}(A^{\text{BF16}}, A^Q) = 10 \log_{10} \left( \frac{\| A^{\text{BF16}} \|_F^2}{\| A^{\text{BF16}} - A^Q \|_F^2} \right) \\[10pt]
    & \text{QSNR}(AB) = 10 \log_{10} \left( \frac{\| A^{\text{BF16}} B^{\text{BF16}} \|_F^2}{\| A^{\text{BF16}} B^{\text{BF16}} - A^Q B^Q \|_F^2} \right)
\end{align}
In this paper, we conduct QSNR analysis on two popular LLMs, Llama 3.1-8B-Instruct and Qwen3-8B. We use the tensors dumped during the inference of each model. We randomly sampled 1000 tensors and use the average value for measuring QSNR.

\subsection{Implications of Fine-Grained Block Quantization: (32 $\to$ 16)}
\label{subsec:bs}
The limited exponent width (E = 2) of FP4 inherently constrains the representable dynamic range to a very small ratio:  $12\times$ ($=6.0/0.5$). 
Consequently, blocks exhibiting high variance inevitably incur increased flush-to-zero (values quantized to zero) rates for smaller magnitude values.
For example, for activation tensors, decreasing the block size from 32 to 16 results in a reduction in flush-to-zero values from 20\% to 13\%, i.e. decreasing the flush-to-zero ratio by 35\%.
While techniques such as using rotation matrices~\cite{mr_gptq} help reduce the dynamic range of the blocks, the relative decrease in flush-to-zero rates remains consistent. Consequently, the reduced block size yields a net increase in QSNR around {\bf{1~dB}}.

\subsection{Impact of Fine-Grained Scaling Factor Format: E8M0 $\to$ E4M3}
While the formats differ in exponent width (4-bit vs. 8-bit), the extended range of the 8-bit exponent in MXFP4 is largely redundant. Notably, for nearly all weight tensors and over 98\% of activation tensors, a 4-bit exponent suffices to capture the scaling factor's dynamic range. Consequently, the MXFP4 scaling factor format leaves four exponent bits unutilized for most tensors—an inefficiency we propose addressing by truncating the exponent from E8 $\to$ E4 for compact storage.

However, the critical functional distinction lies in the mantissa bits. Given that systematic tensor outliers govern quantization thresholds, their precise resolution is paramount for fidelity~\cite{dettmers2022llmint8, smooth_quant}. 
However, MXFP4 (E8M0) lacks mantissa bits, rigidly constraining scaling factors to powers-of-two. This prevents the accurate representation of outliers falling between intervals; for instance, values between $4.0$ and $6.0$ can incur representation errors of up to $20\%$.
Conversely, the E4M3 scaling format (NVFP4) retains three mantissa bits, enabling finer-grained scaling precision that can better approximate the optimal scale for these critical outliers (within 0.2-0.3dB of storing an FP32 scaling factor). Thus, E4M3 effectively minimizes the error for large-magnitude values for the given budget of 8-bits, thereby significantly boosting the tensor’s QSNR. 
We observed a {\bf{3–4~dB}} improvement attributable solely to increased scaling factor mantissa precision and analyze the specific hardware costs of this fidelity gain next.

\subsection{Impact on HW Cost of Block Size and Scaling Factor Format}
\label{subsec:hw_overhead}

\begin{figure}[t]
\centerline{\includegraphics[width=0.9\columnwidth]{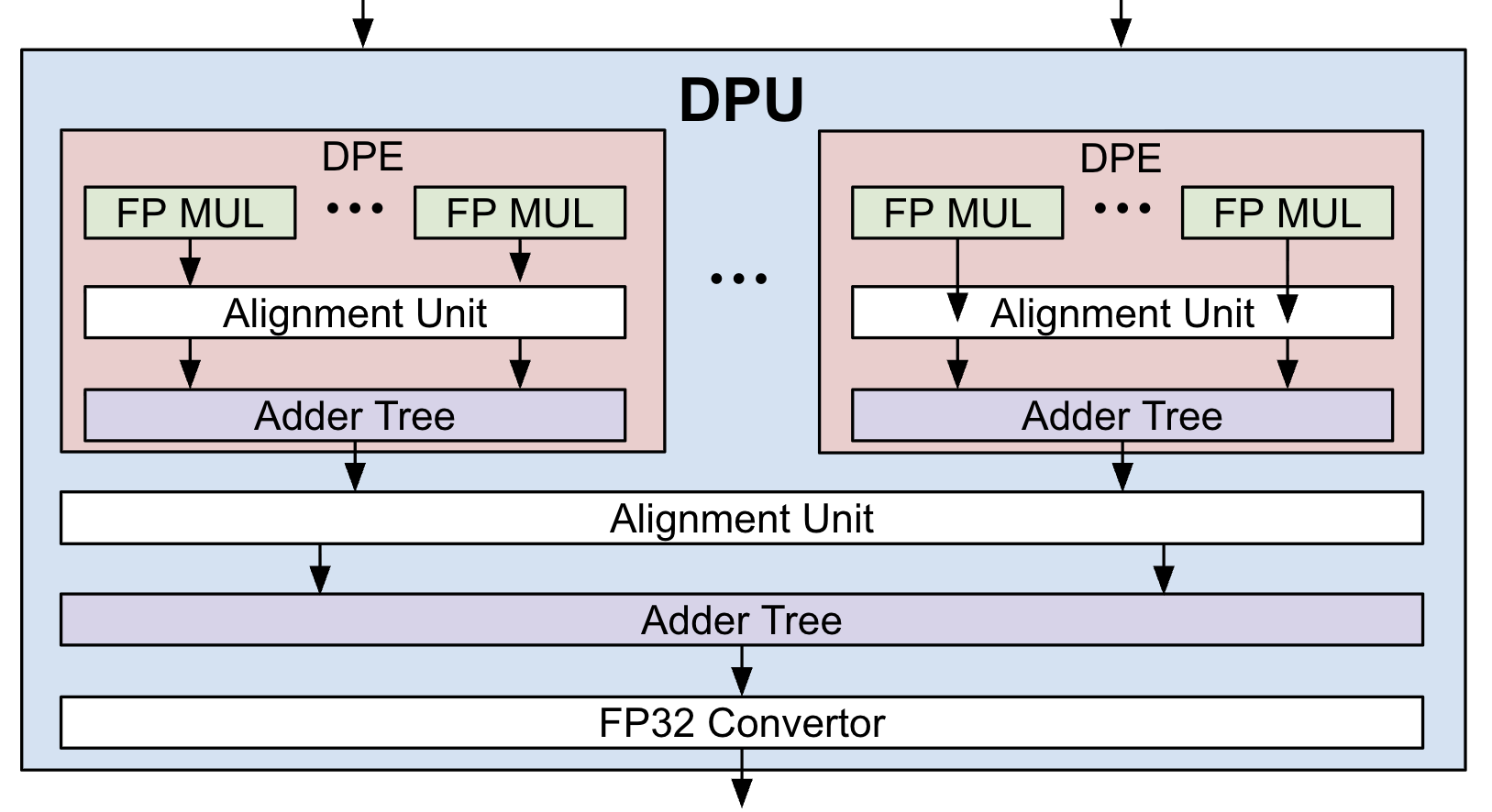}}
\caption{Overview of the DPU architecture.
}
\vspace{-2em}
\label{fig:dpu_hw}
\end{figure}
To understand the HW cost, we extract the baseline area numbers of main components (element-wise matmul, adder trees, shifters and so on) from a production implementation of a multi-format tensor core similar to previous work~\cite{stc2019micro, nervana, 9731711, darvish2023shared} mapped to an advanced TSMC tech node. We use these numbers to build an analytical area model to compare trade-offs between NVFP4 and MXFP4, specifically the impact of (i) the scale factor block size (32 vs. 16) and (ii) the scale factor format (E8M0 vs. E4M3). For a fair comparison, we assume identical hardware tile sizes across designs for tensor cores. 
Tensor core area can be attributed to 2 major components: memory and compute logic as shown in \autoref{fig:tc_hw}.

\textbf{Block size overhead.} We first measure the area impact of using a scale-factor block size of 16 instead of 32. To be conservative, we assume E8M0 scale factors. Based on our area model, a block size of 16 increases tensor-core area by 2\% relative to a block size of 32. This increase is mainly attributed to slightly higher size of SRAM (4.5 bits per element for 16 block size compared to 4.25 bits per element for block size 32) needed for A/B memories and increase in inter block adder tree width.

\textbf{Scale factor format overhead (E4M3 vs. E8M0).} Next, fixing the block size to 16, we quantify the area overhead of using E4M3 rather than E8M0 as the scale-factor format.

First, the capacities of memory for A and memory for B are identical in both cases because the block size is fixed. For example, with block size 16, an 8-bit scale factor, and 4-bit data, each block requires $16 \times 4 + 8 = 72$ bits. The capacity of memory for C is also unchanged because the output precision (we assume this as FP32) does not depend on the input format. Therefore, E8M0 and E4M3 do not differ in memory capacity under the same block size.

In terms of compute logic (i.e. DPU in \autoref{fig:dpu_hw}), the main difference arises in the inter-block alignment logic across DPEs (Dot Product Engines), which (1) resolves the effective scale for each block, (2) computes the maximum exponent ($E_{\max}$), and (3) aligns block-level mantissas with respect to $E_{\max}$ while adding the partial sums across blocks.

With floating-point scale factors (E4M3), scale resolution requires floating-point multiplication: the shared scale is applied to values within the block, requiring \textbf{TensorCoreTileSize / BlockSize floating-point multiplications}. Moreover, the maximum exponent must be determined across all elements, i.e., $\text{NumBlocks} \times \text{BlockSize}$ values, rather than across only NumBlocks values.

With power-of-two scale factors (E8M0), scale resolution requires only one integer addition: the shared scale exponent is added to the block maximum exponent to obtain the effective exponent for the block ($E_{\max,i}$). \textbf{This requires one integer addition per block.} The global maximum exponent is then computed by comparing one value per block (thus total of number of blocks).

As a result, inter-block alignment is substantially more expensive for E4M3 than for E8M0. Based on our area model, using E4M3 as the scale-factor format incurs a 21.3\% compute logic area overhead, 12.6\% total tensor core area overhead relative to E8M0 at the same block size.

\subsection{Proposed Direction}

In summary, we attribute the fidelity gap to fundamental granularity differences between NVFP4 and MXFP4 across two axes: (1) block size and (2) scale factor format. While both improve fidelity, our analysis reveals a stark trade-off: fine-grained scale factors incur high hardware costs, whereas reducing block size is inexpensive.

Consequently, we adopt a finer block size of 16 to leverage spatial locality while retaining the cost-effective coarser E8M0 scale factor format. To recapture the precision of a fine-grained scale factor format without the associated hardware cost, we propose \textit{Overflow-Aware Scaling (OAS)} along with \textit{Macro Block Scaling (MBS)}. This approach allows area-efficient MX hardware to achieve fidelity competitive with NVFP4, effectively decoupling high model performance from expensive hardware requirements.
\section{Enhancing MX Format}
\label{sec:enhancement}
\subsection{Quantization Block Granularity}
As detailed in \autoref{subsec:bs}, reducing the block size is imperative for low-precision formats like FP4. 
While NVIDIA’s hardware restricts MXFP4 to a block size of 32, it natively supports NVFP4 at a finer block size of 16~\cite{cutlass}. We leverage this by utilizing the NVFP4 pipeline but explicitly constraining the block scaling factors to be powers of two (to facilitate the numerical analysis natively). Since the E4M3 format can represent powers of two losslessly (effectively acting as an E4M0 format), this approach allows us to execute MX-style scaling at block size 16 without hardware modification.
This minimal adjustment recovers $1$~dB QSNR. As the majority of scaling factors reside within a $2^{15}$ range of the tensor maximum, the format encapsulates the effective dynamic range with negligible truncation.

\subsection{Overflow-Aware Scaling (OAS)}

Following standard quantization routines, for each $1 \times 16$ block, we compute $\text{SF}_{\text{FP32}} = 6.0 / \alpha_{\max}$ (given FP4 $\text{FP}_{\max} = 6.0$). We obtain the E8M0 scale by masking mantissa bits to enforce the power-of-two constraint. 
The standard computation ensures that $\alpha_{\max}$ maps to the representable range, $(3, 6]$, preventing saturation (clamping) error. 
However, we observe that when $\alpha_{\max} \in [3, 3.5]$ (3.5 being the mid-point of two consecutive representable numbers within 2$\times$ of 6, i.e. 3 and 4), doubling the scaling factor maps the absmax to $[6, 7]$, resulting in saturation since the format limit is 6.0. 
Nevertheless, this shift \textbf{preserves the relative quantization error} for $\alpha_{\max}$ (e.g., quantizing $3.3 \mapsto 3.0$ versus $6.6 \mapsto 6.0$ yields identical relative error).
More broadly, this scaling adjustment maintains relative error fidelity for any block element that previously mapped to the standard FP4 normal range of $[1, 6]$. 
In addition, the key advantage of this approach is that \textbf{it doubles the representable dynamic range} to accommodate lower-magnitude elements, thereby reducing quantization error for the tail of the distribution. 
We call this our Overflow-Aware Scaling (OAS), which applies scaling to map $\alpha_{\max}$ to $(3.5, 7]$.
It is also worthwhile to note that MXFP4-OCP maps $\alpha_{\max}$ to $(4, 8]$, which allows some overflow, but not ideal unlike OAS. For example, if $\alpha_{\max}$ is mapped to 7.6, the quantization error would be $|\frac{(7.6-6)}{7.6}|=21\%$, but if it was mapped to 3.8, then the error could have been $|\frac{(3.8-4)}{3.8}|=5.3\%$.

We implement OAS by checking mantissa bits, and we observe around 15\% of the blocks take advantage of this OAS with no performance overhead compared to MXFP4-OCP while increasing the QSNR by 0.5 dB and largely improving downstream evaluation results as shown in \autoref{sec:eval}.

\subsection{Macro Block Scaling (MBS)}
\label{subsec:mbs}
Outliers play a disproportionate role in quantization fidelity, despite comprising a negligible fraction (typically less than 1\%) of the tensor~\cite{Guo_2023, dettmers2023spqrsparsequantizedrepresentationnearlossless}. A fundamental limitation of the E8M0 scaling format is that its quantization error is strictly a function of the original value, meaning the format lacks the flexibility to prioritize or ``attend" to these critical outlier regions, irrespective of the scaling factor used as it does not change mantissa bits.

To address this, we propose coarser scaling (specifically targeting a $1 \times 128$ block size) with higher precision (with 8 bits of mantissa) as the \textbf{Macro Block Scaling (MBS)}.
Although this granularity is coarser than the fundamental compute block size of $1 \times 16$, we identify $1 \times 128$ as the optimal compromise as shown in \autoref{app:mbs_abl}: it is sufficiently fine-grained to isolate high-magnitude outliers effectively with  extra mantissa bits, yet coarse enough to minimize post-processing overhead and storage costs\footnote{Spatially clustered outliers could potentially benefit from pre-processing techniques such as column reordering~\cite{MLSYS2024_atom}; however, integration of such permutation-based optimizations remains beyond the scope of this work.}.
It is important to note that while both NVFP4 and our proposed MBS approach differ in the approach to store/process scaling factors, our method achieves a critical advantage: it effectively isolates outliers with MBS to preserve model fidelity without the prohibitive hardware cost associated with native fine-grained scaling format (i.e. E4M3 for local scale factor). 
In addition, our quantization strategy maintains minimal computational overhead by eliminating the need for a two-pass traversal over the tensor for scale computation and subsequent quantization. \textbf{Our MBS scheme operates on local $1\times128$ blocks}, which constitutes a natural extension of the fundamental $1\times16$ granularity and is handled seamlessly within the existing CUDA parallelization framework.

\subsubsection{Computation of MBS Factor}
We define the macro-block maximum as $\alpha_{\max}^{128} = \max(\alpha_1^{16}, \dots, \alpha_8^{16})$, where $\alpha_i^{16}$ denotes the absmax within the $i$-th contiguous $1 \times 16$ sub-block.
We assume an algorithm that maps the input $\alpha_{\max}^{128}$ to a scale factor $\text{SF}_{\text{MBS}}^{128}$ (e.g., $\text{SF}_{\text{MBS}}^{128} = 6.0 / \alpha_{\max}^{128}$).
We express the scaling factor as $\text{SF}_{\text{MBS}}^{128} = 2^e (1 + m_{\text{MBS}})$~\cite{mor2025}. Since the local $1 \times 16$ E8M0 scales are purely exponential, they efficiently subsume the macro-exponent $e$, requiring storage only for the mantissa. Empirically, an 8-bit representation approximates the scale within $0.3\%$.
Consequently, we propose storing only the quantized mantissa, denoted as $m_{\text{MBS}}^8$.
Eventually, we use $(1 + m_{\text{MBS}}^8)$ as the actual MBS Factor so the $1 \le \text{MBS Factor} < 2$. 
Upon computing $(1 + m_{\text{MBS}}^8)$, we scale the elements of each $1 \times 16$ block by this factor. This operation shifts the input distribution into the optimal range before we apply the standard MXFP4 quantization, while leveraging our OAS. 

\begin{figure}[t]
\centerline{\includegraphics[width=0.99\columnwidth]{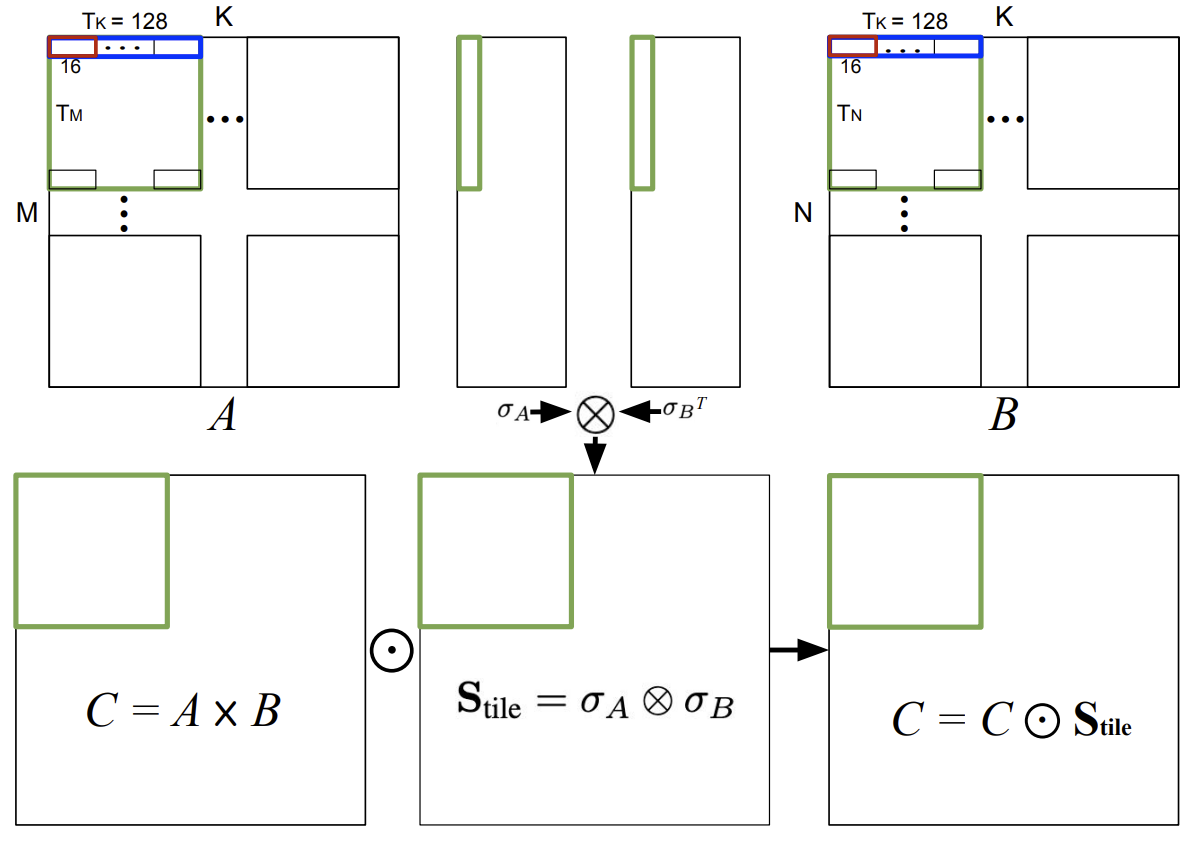}}
\caption{Matrix multiplication $AB^T$ with MBS.
}
\vspace{-1em}
\label{fig:mbs_matmul}
\end{figure}
\subsubsection{Matrix Multiplication with MBS}
 
We perform the matrix multiplication $AB^T$, where $A \in \mathbb{R}^{M \times K}$ and $B \in \mathbb{R}^{N \times K}$. 
The computation adheres to a tiled execution model (e.g., CUTLASS~\cite{cutlass}), wherein discrete tiles of both matrices are iteratively fetched from HBM into the cache hierarchy. We define the tile dimensions as $T_M \times T_K$ for matrix $A$ and $T_N \times T_K$ for matrix $B$. Consequently, the inner kernel executes the product of these tiles, denoted as $(T_M \times T_K) \times (T_N \times T_K)^T$.

We align the kernel to $1 \times 128$ MBS granularity ($T_K = 128$). While the native \texttt{mma.m16n8k64} instruction processes 64-element chunks~\cite{nvidia_ptx_mma_16864}, we leverage CUTLASS to aggregate execution into $128^3$ tiles. This synchronization (\autoref{fig:mbs_matmul}) ensures scaling updates coincide with tile boundaries, enabling efficient epilogue interception without architectural divergence.

At initialization, threads prefetch encoded MBS ($m_{\text{MBS}}^8$) to LLC and compute FP16 scales $\sigma = (1 + m_{\text{MBS}}^8)^{-1}$. The steady-state loop employs a multi-stage pipeline to hide latency, issuing asynchronous instructions (e.g., \texttt{cp.async}) to stage subsequent tiles while concurrently saturating Tensor Cores with the compute-bound FP4 GEMM.
Upon loop termination, the $128 \times 128$ FP32 output tile $\mathbf{C}_{\text{tile}}$ resides in the distributed register file. In the epilogue, we synthesize the de-quantization surface $\mathbf{S}_{\text{tile}} = \sigma_A \otimes \sigma_B$ and fuse it directly into the accumulators via an element-wise Hadamard product: $\mathbf{C}_{ij} \leftarrow \mathbf{C}_{ij} \odot (\sigma_{A,i} \cdot \sigma_{B,j})$. This operation maps to a sequence of low-latency, register-level FP32 FMUL instructions prior to writeback.

From a Roofline perspective (\autoref{app:matmul_overhead}), MBS latency is theoretically hidden provided Vector Core throughput exceeds $\approx 1.56\%$ ($1/64$) of Tensor Core peak. We quantify realized overhead in \autoref{subsec:overhead}. Crucially, we schedule MBS scaling on Vector Cores concurrently with the main workload, leaving Tensor Cores fully dedicated to the dense GEMM. \textit{Significantly, this confirms MBS is strictly a software optimization, requiring no hardware changes.}

\subsubsection{
Optimization for MBS
}
\label{subsec:dynamic_mbs}
\begin{figure}[t]
  \begin{center}
    \centerline{\includegraphics[width=\columnwidth]{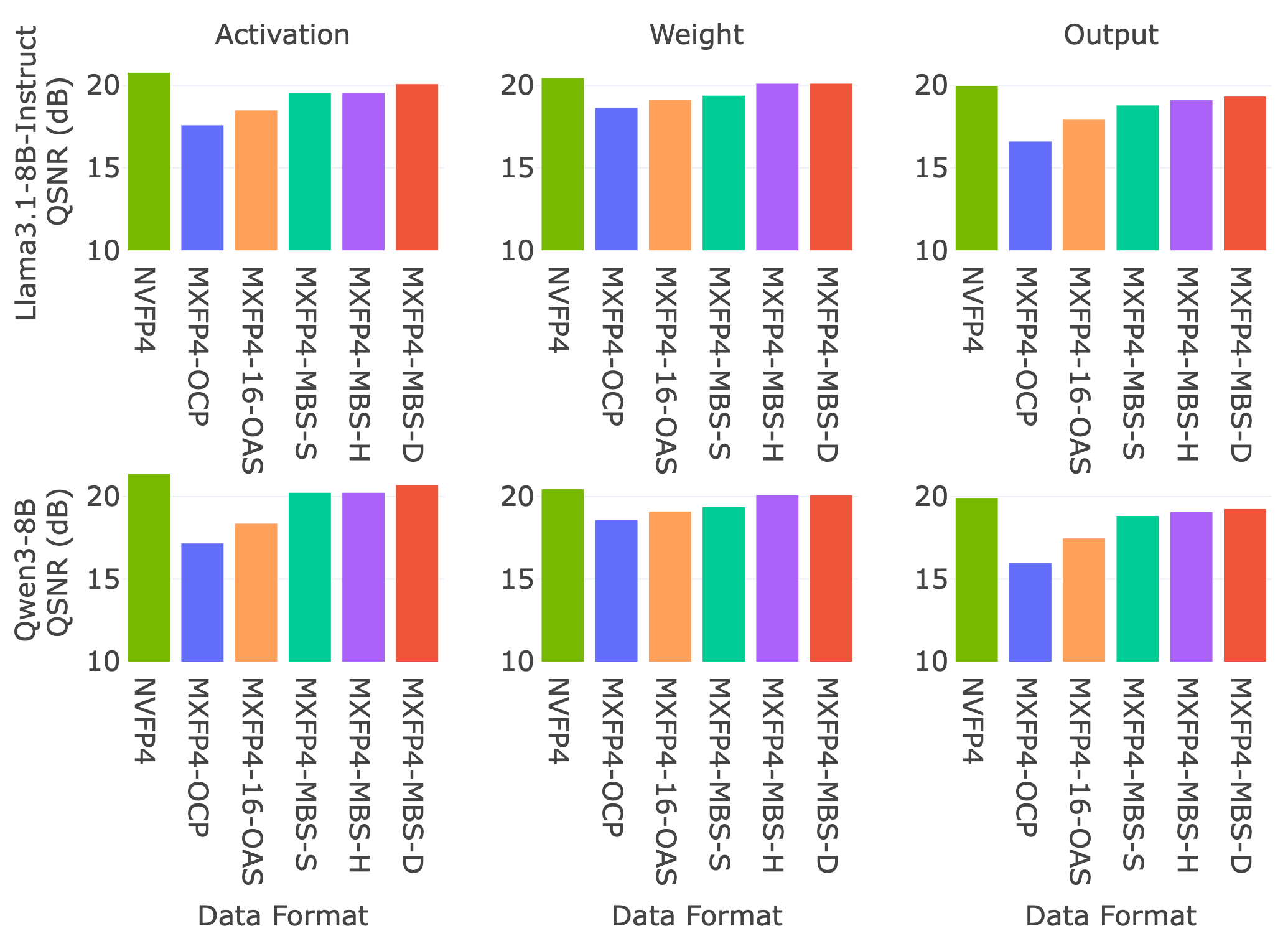}}
    \caption{
      QSNR analysis with different formats on activation, weight, and output of matrix multiplication.
      }
    \label{fig:qsnr}
  \end{center}
  \vspace{-2em}
\end{figure}

For the MBS scheme, 
we compute the scaling factor $(1 + m^8_{\text{MBS}})$ via two proposed algorithms, differentiated by their trade-off between computational overhead and fidelity (QSNR and end-to-end accuracy).

\textit{\textbf{Static:}}
We derive the scaling factor  from the macro-block maximum $\alpha_{\max}^{128}$, computing the reciprocal to normalize to $F_{\max}=6.0$ and extracting the 8 most significant bits:
\begin{equation}
\label{eq:static_mbs}
    m^8_{\text{MBS}} = \left( \text{bits}\left( \frac{6.0}{\alpha_{\max}^{128}} \right) \ \& \ \texttt{0x007F8000} \right) \gg 15
\end{equation}
This operation isolates the target scale's leading mantissa bits, yielding a computationally inexpensive approximation. In terms of fidelity, MBS-Static (\textbf{MBS-S}) improves average QSNR by $+1.1\,\text{dB}$ over the MXFP4 using $1 \times 16$ with OAS.

\textit{\textbf{Dynamic:}} 
While the Static assignment of  $m^8_{\text{MBS}}$ is robust, it lacks MSE guarantees~\cite{mr_gptq}. We address this via a memoization-based search over a narrow range, trading marginal overhead for superior fidelity.

\paragraph{Memoization Strategy} To bypass runtime MSE calculation, we optimize MBS selection via a precomputed Look-Up Table (LUT). For a candidate factor $m_j$ and input $x_i$, we conceptually scale to $x_i \cdot (1 + m_j)$, derive the local scaling factor $SF$ with OAS from the block's maximum magnitude, and define the quantized output as:
\begin{equation}
    \hat{x}_i = Q^{\text{FP4}}\left( x_i \cdot (1 + m_j) \cdot SF \right)
\end{equation}
These tables store squared relative errors, indexed by candidate scale $m_j$ and scaled intermediate value $v_{ij} = x_i \cdot SF$.
\begin{equation}
    \mathcal{T}[v_{ij}, m_j] \approx \left( \frac{\hat{x}_i - x_i}{x_i} \right)^2
\end{equation}
We route sub-normal ($v_{ij} < 1$) and normal ($v_{ij} \ge 1$) values to distinct tables, discretizing the domain into 64 points with 16 slots. The resulting 2,048 entries (4KB in FP16) occupy $<2\%$ of NVIDIA B200 shared memory. Runtime access ensures fully coalesced shared memory reads, with the final factor $m^8_{\text{MBS}} = m_{j^*}$ selected by minimizing macro-block Sum of Squared Errors (SSE):
\begin{equation}
    j^* = \operatorname*{argmin}_{j} \sum_{b=1}^{8} \sum_{i=1}^{16} (x_{bi})^2 \cdot \mathcal{T}[v_{bi,j}, m_j]
\end{equation}
\textit{Overhead Analysis:} 
The search entails one type conversion and one FFMA per slot, costing $\sim32$ ops/element amortized. This fixed per-element cost is negligible relative to the GEMM workload, as it is diluted by the massive, $K$-scaling arithmetic intensity of the core kernel.

\textit{Fidelity Improvement:} 
\textbf{MBS-D} yields a $1.6\,\text{dB}$ QSNR improvement over the MXFP4 using $1 \times 16$ with OAS. We also show (consistent with~\cite{mr_gptq}) that these QSNR gains strongly correlate with the recovery of downstream model accuracy (\autoref{sec:eval})

\subsection{Overall QSNR Comparison of NVFP4 with MX4-MBS-[S/D]}
As shown in \autoref{fig:qsnr}, MBS elevates QSNR from $18.6 \to 20.1\,\text{dB}$ (Weights) and $17.4 \to 19.9\,\text{dB}$ (Activations), narrowing the NVFP4 gap to $<1\,\text{dB}$. This proximity implies \textit{statistically similar errors} and comparable inference convergence--a theoretical equivalence we empirically validate below. Balancing fidelity with runtime cost, we employ {MBS-Dynamic} for Weights and {MBS-Static} for Activations, designating this configuration \textbf{MBS-Hybrid (MBS-H)} as our default for all end-to-end evaluations.
\begin{table*}[t]
  \caption{Downstream evaluation results on \textbf{Llama3.1-8B-Instruct} with different formats.}
  \label{tab:eval_llama}
\vspace{-0.5em}
\begin{center}
    \begin{small}
      \begin{sc}
        \begin{tabular}{lccccccc}
          \toprule
          Precision  & MMLU-Pro         & GSM8K      & Hellaswag & Winogrande & Arc-C & Arc-E & \textbf{Average} \\
          \midrule
          BF16    & 44.22 & 83.18 & 80.07 & 78.61 & 55.29 & 81.82 & 70.53  \\
          \midrule
          MXFP4-OCP & 32.50  & 65.37 & 74.41 & 71.19 & 47.61 & 76.43 & 61.25\\
          MX+ ~\cite{mxplus2025lee} & 34.96 & 69.74 & 74.88 & 71.51 & 47.18	& 77.36 & \textbf{62.61} \\
          \midrule
          MXFP4-16    & 31.34 & 62.49 & 74.02 & 72.14 & 47.70 & 75.63 & 60.55           \\
          MXFP4-16-OAS    & 35.02 & 69.50 & 75.99 & 71.27 & 50.68 & 77.90 & 63.39 \\
          MXFP4-MBS-S & 37.98 & 74.37 & 77.16 & 73.16  & 52.22 & 79.76 & 65.77  \\

          MXFP4-MBS-H & 37.35 & 78.52 & 77.32 & 74.98 & 51.54 & 79.29 & \textbf{66.50} \\
          \midrule
          NVFP4   & 38.83 & 77.12 & 78.66 & 75.69 & 52.05 & 79.76 & \textbf{67.02}\\
          \bottomrule
        \end{tabular}
      \end{sc}
    \end{small}
  \end{center}
\end{table*}

\begin{table*}[t]
  \caption{Downstream evaluation results on 
  \textbf{Qwen3-8B} with different formats.}
  \label{tab:eval_qwen}
\vspace{-0.5em}
\begin{center}
    \begin{small}
      \begin{sc}
        \begin{tabular}{lccccccc}
          \toprule
          Precision  & MMLU-Pro         & GSM8K      & Hellaswag & Winogrande & Arc-C & Arc-E & \textbf{Average}  \\
          \midrule
          BF16    & 63.14 & 90.38 & 76.51 & 70.56 & 56.91 & 83.33 & 73.47  \\
          \midrule
          MXFP4-OCP    & 43.78 & 83.83 & 70.98 & 67.01  & 50.77 & 76.64 & 65.50 \\
          MX+~\cite{mxplus2025lee} & 51.80 & 86.29 & 72.27 & 68.03 & 49.91 & 77.61 & \textbf{67.65} \\
          \midrule
          MXFP4-16    & 49.58 & 83.12 & 71.17 & 68.90 & 48.89 & 79.25 & 66.82         \\
          MXFP4-16-OAS    & 57.85 & 87.52 & 73.14 & 68.03 & 52.56 & 79.17 & 69.71 \\
          MXFP4-MBS-S      & 58.81 & 87.84 & 73.66 & 68.98 & 52.39 & 81.31 & 70.50 \\
          MXFP4-MBS-H & 59.30 & 87.92 & 74.12 & 70.01 & 52.65 & 81.06 & \textbf{70.84} \\
          \midrule
          NVFP4   & 60.94 & 88.78 & 74.66 & 68.43 & 55.03 & 81.06 & \textbf{71.48}\\
          \bottomrule
        \end{tabular}
      \end{sc}
    \end{small}
  \end{center}
  \vskip -0.1in
\end{table*}
\section{Evaluation}
\label{sec:eval}

\subsection{Setups}
To evaluate the proposed enhanced MX formats, we use vLLM~\cite{kwon2023efficient} as the inference engine and use Language Model Evaluation Harness~\cite{eval-harness} as the evaluation framework.
We use Llama 3.1-8B~\cite{grattafiori2024llama3herdmodels}, Qwen3-8B~\cite{qwen3_8b}, Llama 4-Maverick~\cite{llama4_maverick_17b}, and DeepSeek-R1~\cite{deepseek_r1}.
We quantize all linear layers (i.e. QKVO projections, ones in FFN, and each expert in MoE layers). We apply quantization to both weight and activation so that it can actually utilize compute units with reduced precision.
To focus on the effectiveness of the format, we use direct-cast without using any calibration data following the methods used in the previous works~\cite{darvish2023shared, mxplus2025lee}, so our evaluation \textbf{does not} use any re-training and fine-tuning.

\subsection{Benchmarking on Various LLMs}

In \autoref{tab:eval_llama} and \autoref{tab:eval_qwen}, we report downstream evaluation results for different quantization schemes on Llama 3.1-8B-Instruct (L3.1-8B) and Qwen3-8B (Q3-8B).
With MXFP4-OCP, L3.1-8B and Q3-8B achieve average accuracies of 61.25\% and 65.50\% across all benchmarks, respectively.
MX+~\cite{mxplus2025lee}, a state-of-the-art MX scheme that repurposes exponent bits in the per-block maximum, improves average accuracy by 1.76\% over the MXFP4-OCP baseline.
Our MXFP4-16-OAS further improves over MX+ on both L3.1-8B and Q3-8B, with a 1.42\% average gain.
Building on OAS, applying Macro Block Scaling with the static variant (MBS-S) to both activations and weights yields an additional 1.55\% improvement, primarily thanks to better preservation of outliers.
Finally, MXFP4-MBS-H (MBS-S for activations and MBS-D for weights) further improves accuracy by 0.54\%, reducing the remaining gap to NVFP4 to within 1\% on average.

\begin{table}[t]
  \caption{Downstream evaluation results on DeepSeek-R1 with different formats.}
\vspace{-0.5em}
  \label{tab:DS-R1_eval}
  \begin{center}
    \begin{small}
      \begin{sc}
        \begin{tabular}{lccr}
          \toprule
          Precision  & MMLU-Pro & GSM8K \\
          \midrule
          BF16                       & 83.19 & 95.98 &   \\
          \midrule
          MXFP4-OCP                    & 72.52 & 95.91 &   \\
          MX+~\cite{mxplus2025lee}     & 79.85 & 96.13 &   \\
          \midrule
          MXFP4-16                     & 76.29 & 96.66 &          \\
          MXFP4-16-OAS               & 75.36 & 96.29 &  \\
          MXFP4-MBS-S         & 82.37 & 96.82 &   \\
        MXFP4-MBS-H         & 82.06 & 96.89 &          \\
        \midrule
          NVFP4                      & 82.69 & 96.36 &   \\
          \bottomrule
        \end{tabular}
      \end{sc}
    \end{small}
  \end{center}
  \vspace{-1em}
\end{table}

Next, we evaluate our methods on frontier MoE models, including DeepSeek-R1 and Llama 4-Maverick.
Consistent with prior observations~\cite{mr_gptq}, larger models can be less sensitive to quantization; nevertheless, we observe substantial degradation with MXFP4-OCP (up to 10\% on MMLU-Pro for DeepSeek-R1; \autoref{tab:DS-R1_eval}).
Across these models, OAS and MBS substantially recover accuracy, bringing MXFP4 close to (and in some cases on par with) NVFP4.
Please refer to \autoref{app:l4_eval} for Llama 4-Maverick results.
We also report perplexity on Wikitext~\cite{merity2016pointer} in \autoref{app:ppl}, which exhibits the same trend as downstream evaluations and further supports the effectiveness of OAS and MBS.
Overall, these results validate that improved representation fidelity from OAS and MBS translates to consistent end-to-end gains.
\subsection{Overhead Analysis}
\label{subsec:overhead}
For \textbf{MBS-Static (MBS-S)}, we derive $m^8_{\text{MBS}}$ via \autoref{eq:static_mbs} and scale each $\alpha_i^{16}$ by $(1 + m^8_{\text{MBS}})$ to obtain the optimized scaling factor $SF_i$ for the $i$-th sub-block.
We develop a CUDA kernel to execute this logic and quantify the instruction overhead using NVIDIA Nsight Compute~\cite{nsight}. Specifically, while the baseline implementation requires approximately $16.1$ ops per element, our Static-MBS approach introduces a marginal average overhead of only $2.7$ ops/element. Even with this slight arithmetic increase, the kernel remains strictly bound by data access latency, allowing the additional computation to be effectively hidden behind memory latency. 
 Hence, in our proposed MBS-Static, we observe \textbf{zero effective overhead for Activations} which are quantized on-the-fly for MXFP4-MBS-S and MXFP4-MBS-H.

For \textbf{MBS-Dynamic (MBS-D)}, we perform an exhaustive search using 16 potential candidates for the $m^8_{\text{MBS}}$ that reduces the SSE of the $1 \times 128$ block as shown in \autoref{subsec:dynamic_mbs}. In practice, we observe our CUDA implementation to be approximately $2.5\times$--$3\times$ slower than the MBS-S counterpart. 
We believe it can be further optimized, but for the fair conservative evaluation, we use MBS-H (Hybrid), using MBS-D quantization only for Weights while using MBS-S for activations.
This makes sure there is no overhead due to MBS-D during inference as activation still uses MBS-S.

For the actual matrix multiplication with MBS, we implement the GEMM kernel with the MXFP4-MBS-H using CUTLASS 4.3.0 on the NVIDIA Blackwell (SM100) architecture, building upon the MXFP4-OCP GEMM kernel which utilizes E2M1 data with E8M0 scale factors at a $1\times32$ granularity. Our implementation extends the CUTLASS warp-specialized mainloop via a custom dispatch policy that intercepts the loop at an MBS block size of 128. Computation is distributed across a cluster of four Cooperative Thread Arrays (CTAs) configured as $2\times2\times1$, which cooperatively compute each output tile (e.g., $256\times256$). The cluster is organized into two leader-peer pairs, with each pair processing a $128\times256$ region of the output. In this configuration, leader CTAs execute the tensor core MMA operations, while peer CTAs facilitate TMA multicast data loading to minimize global memory bandwidth consumption. We provide further explanation in \autoref{app:mbs_impl}.

For the decode stage, the overhead is minimal as the execution is memory-bound due to weight loading, consistent with observations in MX+~\cite{mxplus2025lee}.
During prefill, MXFP4-MBS-H incurs a 6.2\% overhead on top of the baseline GEMM kernel on average for different shapes as summarized in \autoref{tab:static_mbs_overhead}, which is substantially lower than the 54\% overhead reported by MX+~\cite{mxplus2025lee}.
For end-to-end execution time, MXFP4-MBS-H adds negligible overhead for LLM inference.

\begin{table}[t]
  \caption{Overhead GEMM kernel between MXFP4-OCP and our MXFP4-MBS-H. Throughput is measured in TFLOPS.}
  \vspace{-1em}
  \label{tab:static_mbs_overhead}
  \begin{center}
    \begin{small}
      \begin{sc}
        \begin{tabular}{lrrr}
          \toprule
          Shape & MXFP4-OCP & MXFP4-MBS-H & Overhead \\
          \midrule
          512   & 35.90   & 30.57   & +14.84\% \\
          1024  & 285.04  & 259.34  & +9.01\% \\
          2048  & 1669.40 & 1666.63 & +0.17\% \\
          4096  & 4462.95 & 4354.87 & +2.42\% \\
          8192  & 4745.00 & 4520.53 & +4.73\% \\
          \bottomrule
        \end{tabular}
      \end{sc}
    \end{small}
  \end{center}
  \vspace{-2em}
\end{table}
\section{Related Work}
\label{sec:related}
Prior work explores improving block-based low-precision formats. BDR introduces short microexponents and motivates MX-style formats \cite{darvish2023shared}. Several methods treat outliers specially—by reallocating precision from nearby “victim” values \cite{Guo_2023}, using structured sparsity to mix precisions efficiently \cite{jeong2024sdq}, or adding interconnect support for heterogeneous bit-widths \cite{microscorpiq}. However, most require hardware changes, limiting deployment on commodity GPUs.
MX+ is the closest prior work \cite{mxplus2025lee}: it stores extra mantissa bits for the block maximum on top of MXFP4-OCP and runs on GPUs via on-the-fly conversion, but it adds an extra sparse GEMM and can incur up to 54\% overhead.
Accuracy can be further improved with calibration and post-training quantization; e.g., GPTQ-style PTQ narrows the gap to the base model \cite{mr_gptq,frantar2023gptqaccurateposttrainingquantization}. Our method could also benefit from PTQ and QAT, which we leave to future work.
\section{Conclusion}
\label{sec:conclusion}
In this paper, we analyze the MX format and identify the key factors underlying its accuracy gap relative to NVFP4. Based on these insights, we propose OAS and MBS, simple drop-in techniques that strengthen MXFP4.
With these enhancements, MXFP4 reduces the accuracy loss of standard MXFP4-OCP by 62\% on average, shrinking the gap between MXFP4 and NVFP4 from 10\% to $<$1\% on average. Overall, our results show that MXFP4 can deliver near-parity with NVFP4, enabling efficient and accurate quantization for LLM inference.

\section*{Impact Statement}
This paper presents work whose goal is to advance the field of Machine
Learning. There are many potential societal consequences of our work, none which we feel must be specifically highlighted here.


\bibliography{example_paper}
\bibliographystyle{icml2026}

\newpage
\appendix
\onecolumn
\section{Ablation Study with MBS Block Size}
\label{app:mbs_abl}

\begin{figure}[H]
  \begin{center}
    \centerline{\includegraphics[width=0.7\columnwidth]{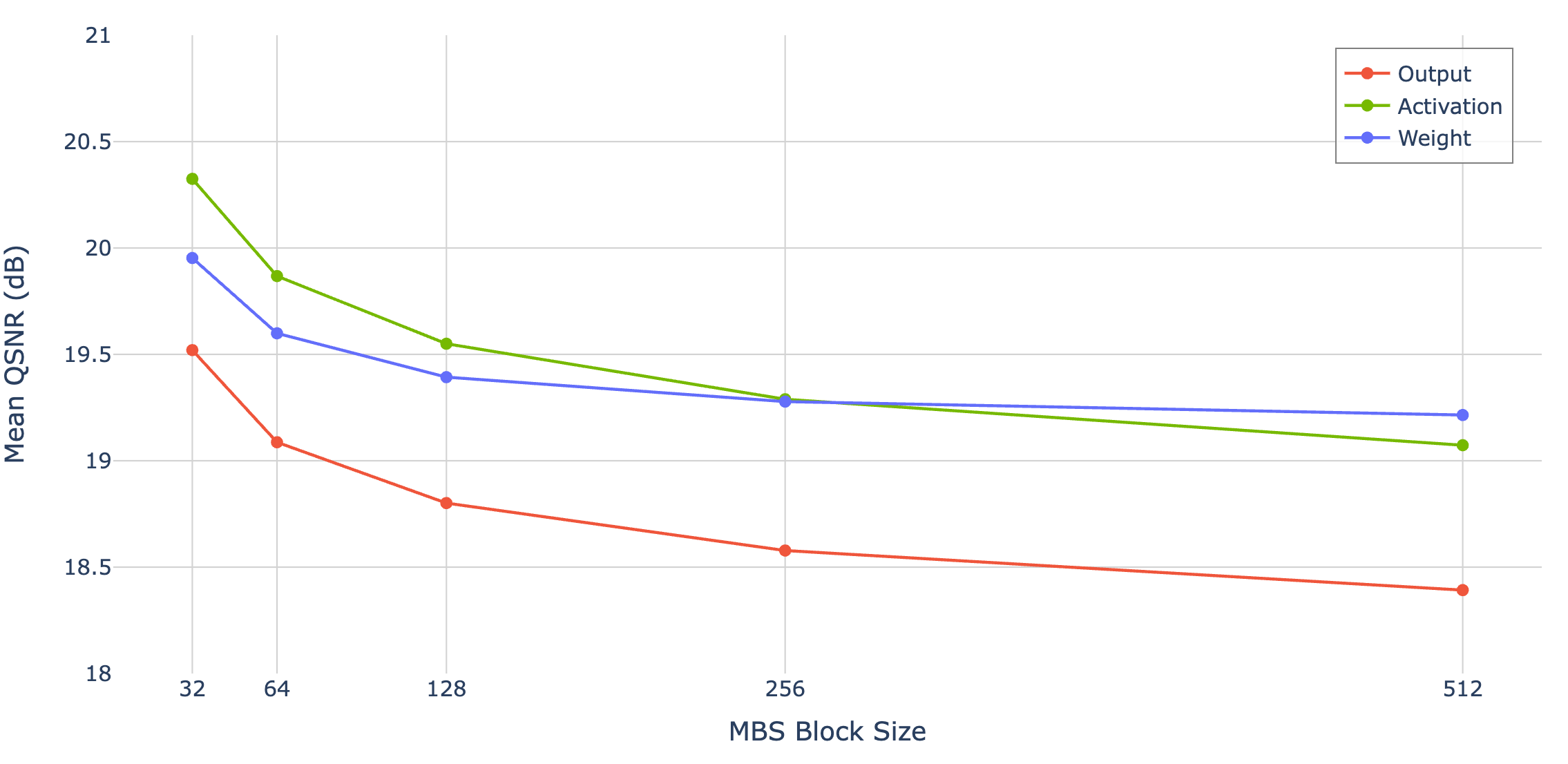}}
    \caption{
      Ablation study for MBS block size with Llama 3.1-8B-Instruct.
      }
    \label{fig:l3_abl}
  \end{center}
  \vspace{-1em}
\end{figure}
\autoref{fig:l3_abl} presents the impact of MBS block size on quantization quality for the Llama3.1-8B-Instruct. We evaluate Mean QSNR across three key components: activation, weight, and output similar to \autoref{fig:qsnr}.

All three metrics show a consistent downward trend as MBS increases, indicating that larger block sizes lead to degraded quantization quality. The total degradation from MBS=32 to MBS=512 is approximately 1.1 dB for output QSNR, 1.2 dB for activation, and 0.7 dB for weight quantization. 

MBS=128 emerges as a favorable operating point, offering a practical balance between quantization quality and hardware efficiency. At this configuration, the model retains 96\% of the output QSNR observed at MBS=32.

\begin{figure}[H]
  \begin{center}
    \centerline{\includegraphics[width=0.7\columnwidth]{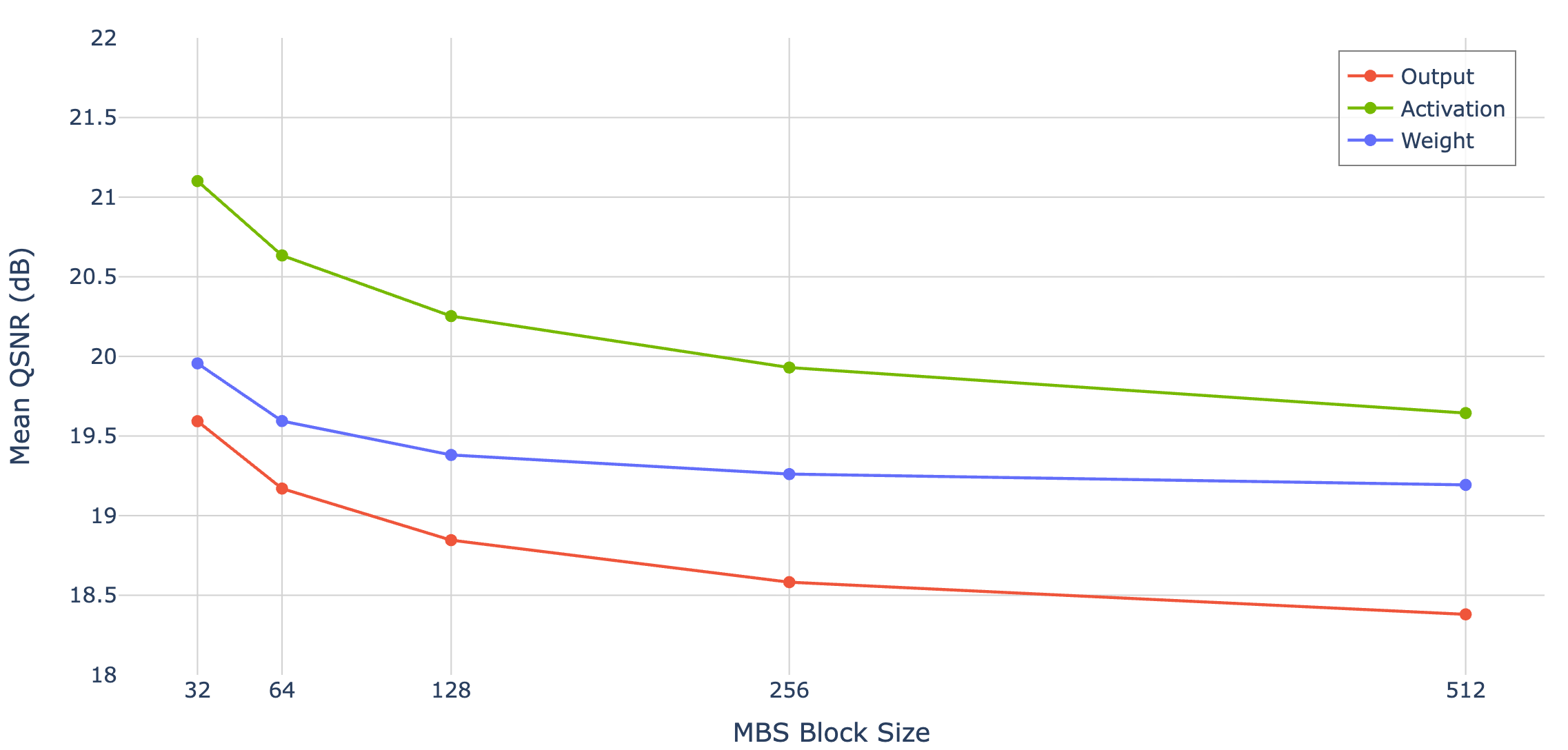}}
    \caption{
      Ablation study for MBS block size with Qwen3-8B.
      }
  \label{fig:qwen3_abl}
  \end{center}
\end{figure}

In \autoref{fig:qwen3_abl}, we show MBS ablation results for the Qwen3-8B.
The model exhibits similar degradation patterns to Llama3.1-8B as MBS increases, suggesting consistent quantization behavior across different model architectures. 
The relatively modest quality degradation of 0.75 dB in output QSNR compared to MBS=32 makes MBS=128 an attractive choice.

\section{Matrix Multiplication Overhead with MBS}
\label{app:matmul_overhead}

\textbf{Overhead Analysis}: We analyze the computational and memory overhead introduced by the MBS scaling mechanism, utilizing the target tile configuration of $T_M \times T_N \times T_K = 128 \times 128 \times 128$.

\textit{Computational Overhead:}
The baseline FP4 tensor operation performs $2 \cdot 128^3$ FP4 FLOPs per tile. Our proposed correction adds $2 \cdot 128^2$ FP32 operations (specifically, vector multiplications). The count ratio of additional FP32 MULs to baseline FP4 MULOPs is:
\begin{equation}
    \frac{\text{Ops}_{\text{MBS (FP32)}}}{\text{Ops}_{\text{TC (FP4)}}} = \frac{2 \cdot 128^2}{1 \cdot 128^3} = \frac{2}{128} \approx 1.56\%
\end{equation}

\textit{Memory Traffic Overhead:}
For the $128 \times 128$ output tile (64 KB), the MBS scheme loads two scaling vectors ($\sigma_A, \sigma_B$) totaling 512 bytes.
\begin{equation}
    \frac{\text{Traffic}_{\text{MBS}}}{\text{Traffic}_{\text{Tile}}} = \frac{512 \text{ bytes}}{65,536 \text{ bytes}} \approx 0.78\%
\end{equation}
This negligible increase in data movement ensures that the kernel's arithmetic intensity remains virtually unchanged.
We also implement the proposed MBS on NVIDIA B200 GPUs and show the analysis in \autoref{subsec:overhead}.

\section{Downstream evaluation results on Llama 4-Maverick }
\label{app:l4_eval}
\begin{table}[H]
  \caption{Downstream evaluation results on Llama 4-Maverick with
different formats.}
  \label{tab:l4_eval}
  \begin{center}
    \begin{small}
      \begin{sc}
        \begin{tabular}{lccr}
          \toprule
          Precision  & MMLU-Pro & GSM8K \\
          \midrule
          BF16            & 80.98 & 94.16 &   \\
          \midrule
          MXFP4-OCP         & 77.73 & 92.42 &   \\
          MX+~\cite{mxplus2025lee}     & 78.46 & 92.80 &   \\
          \midrule
          MXFP4-16          & 78.12 & 92.27 &          \\
          MXFP4-16-OAS    & 78.64 & 93.18 &  \\
          MXFP4-MBS-S           & 79.38 & 94.16 &   \\
        MXFP4-MBS-H           & 79.77 & 93.86 &          \\
        \midrule
          NVFP4          & 80.06 & 94.01 &   \\
          \bottomrule
        \end{tabular}
      \end{sc}
    \end{small}
  \end{center}
\end{table}
In \autoref{tab:l4_eval}, we show the evaluation results using different FP4 formats. 

\section{Perplexity evaluation}
\label{app:ppl}
\begin{table}[H]
  \caption{Word-level perplexity evaluation on Wikitext with Llama3.1-8B-Instruct and Qwen3-8B using different formats.}
  \label{tab:ppl_eval}
  \begin{center}
    \begin{small}
      \begin{sc}
        \begin{tabular}{lccr}
          \toprule
          Precision  & Llama3.1-8B-Instruct & Qwen3-8B \\
          \midrule
          BF16    & 8.82 & 12.20 &   \\
          \midrule
          MXFP4-OCP    & 11.49 & 15.18 &   \\
          MX4+~\cite{mxplus2025lee}    & 10.82 & 14.51 &   \\
          \midrule
          MXFP4-16    & 11.52 & 15.15 &          \\
          MXFP4-16-OAS    & 10.57 & 13.65 &  \\
          MXFP4-MBS-S & 10.04 & 13.09 &   \\
        MXFP4-MBS-H & 9.88 & 13.03 &          \\
        \midrule
          NVFP4   & 9.68 & 12.69 &   \\
          \bottomrule
        \end{tabular}
      \end{sc}
    \end{small}
  \end{center}
\end{table}
In ~\autoref{tab:ppl_eval}, we show the perplexity evaluation results using our methods. Using OAS and MBS, we can reduce the perplexity gap against NVFP4 from 1.82 to 0.20 for Llama3.1-8B-Instruct and 2.49 to 0.34 for Qwen3-8B.

\section{Detailed Explanation for MBS Implementation}
\label{app:mbs_impl}
For our MBS implementation, the memory hierarchy spans three levels: global memory (HBM), which stores the FP4 tensors, E8M0 scale factors, E0M8 MBS scales, and final output; shared memory (SMEM), which holds multi-stage buffered tiles for software pipelining; and tensor memory (TMEM), which stores the MMA accumulators and scale factor registers. Data movement adheres to the CUTLASS producer-consumer pipeline model: producer threads issue asynchronous Tensor Memory Accelerator (TMA) loads for the subsequent K-tile, while consumer threads execute MMA operations on the current tile, synchronized via pipeline barriers. FP4 data and E8M0 scale factors are loaded via TMA with multicast support; subsequently, scale factors are transferred from SMEM to TMEM using Unified Tensor Copy Protocol (UTCCP) operations for block-scaled MMA execution.

The critical algorithmic modification involves intercepting the MMA inner loop to apply MBS at 128-element boundaries. For each K-tile of 256 elements, the kernel performs block-scaled MMA for the first 128 K-elements, accumulating results into a local TMEM accumulator. Directly applying MBS at this stage would incur significant latency by introducing memory traffic and synchronization barriers into the critical compute path. To mitigate this, we employ a TMEM double-buffering strategy with proper warp specialization. The original implementation has warp 0 to perform the MMA operation, warps 1-3 for scheduling, memory load, and epilogue load, warps 4-7 for epilogue processing. But during the main loop execution, the epilogue warps are idle, so we repurpose them for MBS computations. Once the partial sums for a sub-tile are available in the first TMEM buffer, the MMA warp immediately switches to the second TMEM buffer for the next sub-tile. Simultaneously, the MBS warps retrieves the partial sums from the first TMEM buffer into registers and applies the corresponding MBS scales (loaded directly from global memory, as they are small and accessed sequentially) as described in \autoref{subsec:mbs}.


\end{document}